\documentclass  [12pt,letterpaper]{article}
\pdfoutput=1
\usepackage     {jheppub}
\usepackage     {journals}
\usepackage	{graphicx}
\usepackage     {verbatim}
\usepackage     [mathscr]{eucal}
\usepackage     {bm}
\bibpunct       {[}{]}{,}{n}{}{;}

\def\Nfour	{\mathcal N\,{=}\,4}

\def\Nc		{N_{\rm c}}

\def\suck[#1]#2{\includegraphics[#1]{#2}}        

\title		{Holography and off-center collisions of localized shock waves}
\author[a]	{Paul~M.~Chesler,}
\author[b]	{Laurence~G.~Yaffe}
\affiliation[a]	{Department of Physics, Harvard University, Cambridge, MA 02138, USA}
\affiliation[b]	{Department of Physics, University of Washington, Seattle, WA 98195, USA}
\emailAdd	{pchesler@physics.harvard.edu}
\emailAdd	{yaffe@phys.washington.edu}

\date{\today}

\abstract
    {%
    Using numerical holography, we study the collision,
    at non-zero impact parameter,
    of bounded, localized distributions of energy density 
    chosen to mimic relativistic heavy ion collisions,
    in strongly coupled $\Nfour$ supersymmetric Yang-Mills theory.
    Both longitudinal and transverse dynamics in the dual field
    theory are properly described.
    Using the gravitational description,
    we solve 5D Einstein equations
    with no dimensionality reducing symmetry restrictions
    to find the asymptotically anti-de Sitter spacetime geometry.
    Implications of our results on the understanding of early stages of
    heavy ion collisions,
    including the development of transverse radial flow,
    are discussed.
    }

\arxivnumber    {1501.04644}
\begin		{document}
\maketitle

\section	{Introduction}

The recognition that the quark-gluon plasma (QGP) produced in relativistic
heavy ion collisions is strongly coupled \cite{Shuryak:2004cy} has prompted
much work using gauge/gravity duality (or ``holography'') to study aspects
of non-equilibrium dynamics in strongly coupled $\Nfour$ supersymmetric Yang-Mills
theory (SYM), which may be viewed as a theoretically tractable toy model for real QGP{}.
Work to date has involved various idealizations
(boost invariance, planar shocks, imploding shells) which reduce the
dimensionality of the computational problem, at the cost of eliminating
significant aspects of the real collision dynamics
\cite
    {Chesler:2009cy,
    Chesler:2010bi,
    Casalderrey-Solana:2013aba,
    vanderSchee:2013pia,
    Bantilan:2012vu,
    Bantilan:2014sra}.
In this paper, we report the results of the first calculation of this type
which does not impose unrealistic dimensionality reducing restrictions.
We study the collision, at non-zero impact parameter,
of bounded, localized distributions
of energy density 
which mimic colliding relativistic Lorentz-contracted nuclei and form
stable incoming projectiles in strongly coupled SYM.

In the dual gravitational description our initial state consists of two
localized incoming gravitational waves,
in asymptotically anti-de Sitter (AdS) spacetime,
which are arranged to collide at a non-zero impact parameter.
The precollision geometry contains a trapped surface and the collision 
results in the formation of a black brane.
We numerically solve the full 5D Einstein equations for the 
geometry during and after the collision and report on the evolution of the 
SYM stress-energy tensor $ T^{\mu \nu} $.

\section	{Gravitational formulation}
\subsection	{Single shocks}

We construct initial data for Einstein's equations by combining
the metrics describing gravitational shock waves moving at the speed of light
in opposite directions.
In Fefferman-Graham (FG) coordinates,
the metric of a single shock moving in the $\pm z$ direction
is given by \cite{Gubser:2008pc,Grumiller:2008va}
\begin{align}
    ds^2 &=  \frac{L^2}{s^2}
    \left[
	{-} dt^2 + d\bm x_\perp^2 + dz^2 + ds^2
	+ h_\pm(\bm x_\perp, z_\mp,s) \, dz_\mp^2
    \right] \,,
\label{eq:FG}
\end{align}
(where $\bm x_\perp \equiv \{x,y\}$ and $z_\mp \equiv z \mp t$).
This is a sourceless solution to Einstein's equations with cosmological constant
$\Lambda \equiv -6/L^2$,
provided the function $h_\pm$ satisfies the linear
differential equation
\begin{equation}
    \left(\partial_s^2 - \tfrac 3s \, \partial_s + {\bf \nabla}_\perp^2 \right) h_\pm = 0 \,.
\end{equation}
Solutions to this equation which vanish at the boundary, $s=0$,
may be written in the form
\begin{align}
    h_\pm({\bm x}_\perp,z_\mp,s) &=
    \int \frac{d^2{\bf k}_\perp}{(2\pi)^2} \>
    e^{i {\bf k}_\perp \cdot {\bf y}_\perp} \,
    \widetilde H_\pm({\bf k}_\perp,z_\mp) \,
    8 (s^2/k_\perp^2) \, I_2(k_\perp s) \,,
\label{eq:h}
\end{align}
where $\widetilde H_\pm$ is an arbitrary function of the
2D transverse wavevector $\bm k_\perp$ and the longitudinal variable $z_\mp$
(and $I_2$ is a modified Bessel function).

The geometry described by eqs.~(\ref{eq:FG}) and (\ref{eq:h})
represents a state in the dual SYM theory with a stress-energy
tensor expectation value given by \cite{deHaro:2000xn}.
\begin{equation}
\label{eq:singleshock}
    \langle T^{00} \rangle =
    \langle T^{zz} \rangle =
    \pm \langle T^{0z} \rangle = \kappa \, H_\pm(\bm x_\perp,z_\mp),
\end{equation}
with all other components vanishing.
Here, $H_\pm$ is the 2D transverse Fourier transform of $\widetilde H_\pm$,
and the constant $\kappa \equiv L^3/(4\pi G_N) = \Nc^2 / (2\pi^2)$
(with $\Nc$ the gauge group rank of the $SU(\Nc)$ SYM theory).
In other words, the function $H_\pm$ specifies the energy density
(and longitudinal stress and momentum density) of a shock wave moving,
non-dispersively, at the speed of light in the $\pm z$ direction.
Given any choice of this energy density profile,
eqs.~(\ref{eq:FG}) and (\ref{eq:h}) give an explicit form for the
unique dual gravitational geometry which describes this shock wave.

For simplicity, we consider Gaussian energy density profiles,
\begin{equation}
\label{eq:initialH}
    H_\pm(\bm x_\perp,z_\mp)
    =
    \frac{\mathcal A}{\sqrt{2 \pi w^2}} \,
    \exp \left({-\tfrac 12 z_\mp^2/w^2 }\right) \,
    \exp \left[{-\tfrac 12 (\bm x_\perp \mp \bm b/2)^2/R^2}\right] \,,
\end{equation}
with longitudinal width $w$, transverse width $R$,
and a transverse offset $\pm \bm b/2$.
Hence, $\bm b$ will be the impact parameter
when these oppositely directed shock waves collide.
The amplitude $\mathcal A$ gives the maximum value of the longitudinally integrated
energy density (divided by $\kappa$).

For the explicit computations presented below,
we adopt units in which the amplitude $\mathcal A$
equals unity and choose 
longitudinal and transverse widths $w = \frac 12$ and $R = 4$, respectively,
and impact parameter $\bm b = \frac 34 R \, \hat {\bm x}$.
Consequently, the stress tensor (\ref{eq:singleshock}) describes localized lumps of energy 
centered about $x = \pm b/2$, $y = 0$, and $z = \pm t$.
The AdS curvature scale $L$ is not a physical scale in the dual QFT and may
independently be set to unity.

\subsection {Infalling coordinates}

Our time evolution scheme \cite{Chesler:2013lia} for asymptotically-AdS
gravitational dynamics uses infalling Eddington-Finkelstein (EF) coordinates
in which the spacetime metric has the form
\begin{equation}
    ds^2 = \frac{r^2}{L^2} \, g_{\mu \nu}(x,r) \, dx^\mu dx^\nu + 2 \, dr \, dt \,,
\label{eq:ansatz}
\end{equation}
with
Greek indices denoting spacetime boundary coordinates, $x^\mu \equiv (t,x,y,z)$,
and $r$ the bulk radial coordinate.
For the asymptotically-AdS geometries of interest,
the metric coefficients $g_{\mu\nu}$ have the near-boundary behavior
$g_{\mu \nu} = \eta_{\mu \nu} + g_{\mu \nu}^{(4)}/r^4 + O(1/r^5)$
as $r \to \infty$,
with $\|\eta_{\mu\nu} \| \equiv \mathrm{diag}(-1,+1,+1,+1)$
the usual Minkowski metric tensor.
The sub-leading coefficients $g^{(4)}_{\mu\nu}$ determine 
the SYM stress tensor expectation value.
In these coordinates,
\begin{equation}
\label{eq:holostress}
    \langle T_{\mu \nu}\rangle/\kappa
    = g_{\mu \nu}^{(4)}  + \tfrac{1}{4} \, \eta_{\mu \nu}\, g_{00}^{(4)}.
\end{equation}

Although the single shock solution to Einstein's equations has the nice analytic form
(\ref{eq:singleshock}) in Fefferman-Graham coordinates,
this geometry does not have a simple analytic form in infalling EF coordinates;
the transformation to infalling coordinates must be performed numerically.
One may easily show that, in infalling coordinates, curves along which $r$
varies, with the other coordinates held fixed, are infalling null geodesics
(with $r$ an affine parameter).
To transform the geometry (\ref{eq:FG}) to the infalling form (\ref{eq:ansatz})
one must locate the same congruence of infalling radial null geodesics in FG coordinates.
Let $Y \equiv \{y^\mu,s\}$ denote the FG coordinates of some event,
and let $X \equiv \{ x^\mu, r\}$ denote the EF coordinates of the event
at affine parameter $r$ along the radial infalling geodesic which begins
at boundary coordinates $x^\mu$.
Then the solution $Y(X)$ to the geodesic equation (in FG coordinates)
for the same null geodesic which begins at boundary coordinates $x^\mu$ 
provides the required mapping between EF and FG coordinates.
Given this mapping,
the required transformation of the FG metric components
$\widetilde G_{MN}(Y)$ to the metric components $G_{MN}(X)$
in our infalling coordinates
(with $ds^2 = G_{MN}(X) \, dX^M  dX^N = \widetilde G_{MN}(Y) \, dY^M dY^N$)
is simply
\begin{equation}
    G_{MN}(X) =
    \frac{\partial Y^A}{\partial X^M} \,
    \frac{\partial Y^B}{\partial X^N} \,
    \widetilde G_{AB}(Y(X)) \,.
\end{equation}

To compute the congruence $Y(X)$, we first
periodically compactified spatial directions
(to obtain a finite computational domain)
with transverse size $L_x = L_y  = 32$ and longitudinal length $L_z = 12$.
We employed spectral methods \cite{Chesler:2013lia}, and used a rectangular grid
built from single domain Fourier grids with 32 points in the transverse directions,
a 256 point Fourier grid in the longitudinal direction, and
three Chebyshev domains of 32 points each in the radial direction.
We used a Newton iterative procedure to solve the non-linear geodesic equation
starting at each grid point on the boundary (modulo the $C_{4v}$ transverse cubic symmetry).%
\footnote
    {%
    To obtain solutions with high accuracy,
    we used 40 digit arithmetic in this step.
    This allowed us to reduce both numerical arithmetic and iterative convergence errors to
    negligible levels, leaving the spectral truncation error as the limiting numerical issue.
    }
We integrated the geodesic equations to a depth of $s = 5$ and fixed the residual
radial shift reparameterization freedom \cite{Chesler:2013lia} by demanding that
the surfaces $u \equiv 1/r = 5$ in infalling coordinates, and $s = 5$ in FG coordinates,
coincide.

\subsection {Colliding shocks: initial data}

For early times, $t \ll -w$, the Gaussian profiles 
$H_\pm$ of the oppositely directed incoming shocks
have negligible overlap and the precollision geometry can
be constructed by replacing the last term in the single shock metric
(\ref{eq:FG}) with the sum of corresponding terms from
left and right moving shocks.
The resulting metric satisfies Einstein's equations, at early times,
up to exponentially small errors.

Initial data for the subsequent time evolution consists of the values
of the spatial metric coefficients, rescaled to have unit determinant,
$
    \hat g_{ij} \equiv g_{ij} / (\det \| g_{ij} \|)^{1/3}
$,
on every point of the initial time slice,
together with the values of the energy and momentum density 
(or equivalently the asymptotic coefficients $g^{(4)}_{0\nu}$)
at the initial time.
The initial time slice was chosen to lie at $t = -2$,
We slightly modified the initial data by adding a small
uniform background energy density,
equal to 3.7\% of the peak energy density of the incoming shocks. 
This modestly displaced the location of the apparent horizon
toward the boundary,
improving numerical stability and allowing use of a coarser grid,
thereby reducing memory requirements.%
\footnote
    {%
    Note added in proof: we have been able to recompute the evolution
    using a much smaller background energy density, equal to 0.05\%
    of the peak energy density.
    Over the time duration studied, the impact on the resulting
    dynamics of this change in background energy density is negligible.
    None of the features of the results discussed below are significantly
    affected by this background energy density.
    }

\subsection	{Colliding shocks: time evolution}

Time development of the geometry was calculated using the procedure described
in detail in ref.~\cite{Chesler:2013lia}.
Time evolution was performed using a spectral grid of size $N_x = N_y = 39$,
$N_z = 145$, and $N_r = 40$.
A fourth-order Runge-Kutta time integration algorithm was used with a time step
of 0.005.  Integration continued to a final time of $t = 4$.
Even though our gravitational dynamics is
five dimensional, with no symmetry restrictions imposed,
we were able to perform the time evolution on a 6 core desktop computer 
with a 14 day runtime.

\section	{Results}

\begin{figure*}[ht]
\suck[scale = 0.50]{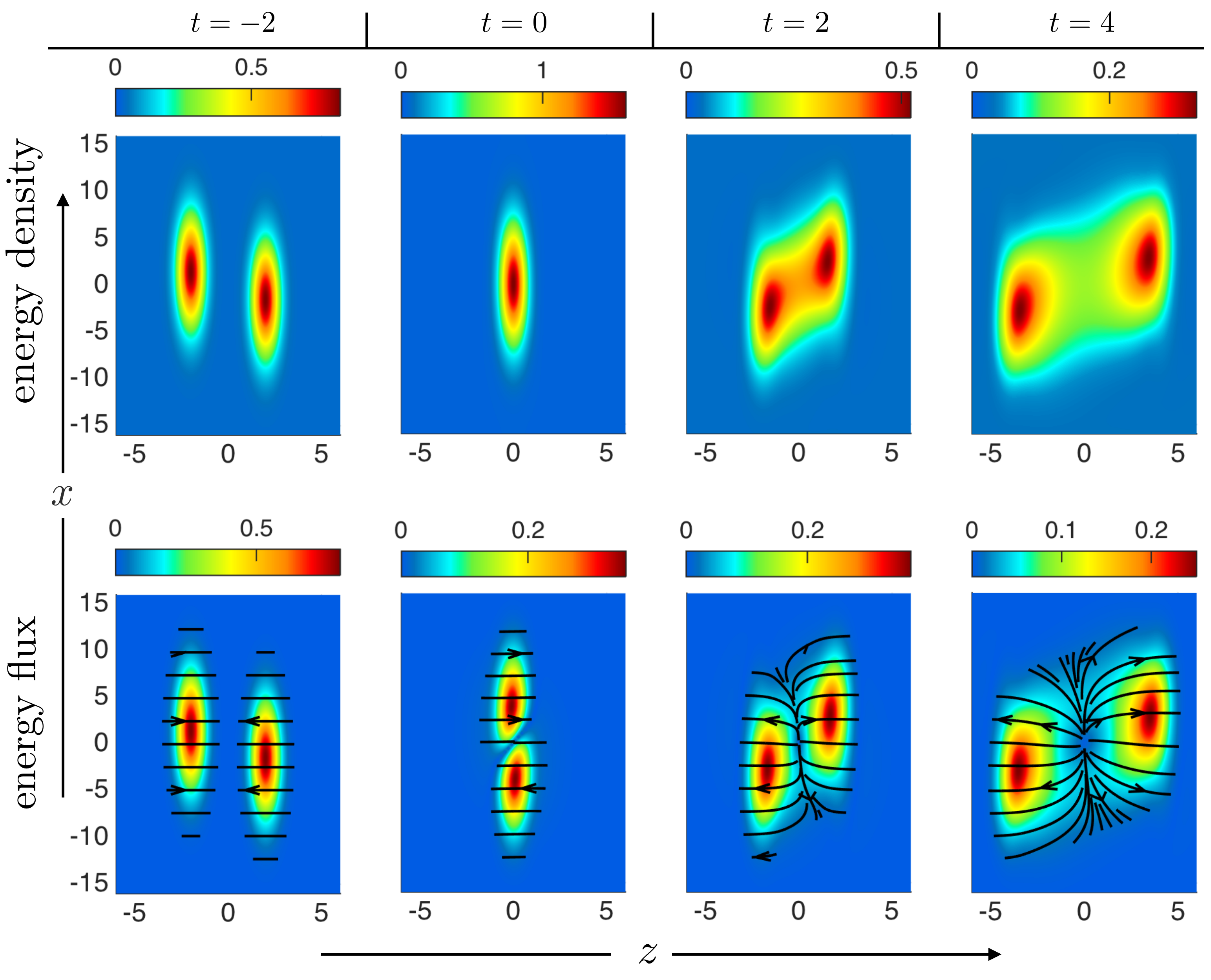}
\caption
    {%
    The energy density $T^{00}$ (top) and
    energy flux $|T^{0i}|$ (bottom),
    at four different times, in the plane $y = 0$.  
    Streamlines in the lower plots
    denote the direction of energy flux.  
    Note that the color scaling varies from plot to plot.
    At the initial time $t = -2$ the shocks are at $z = \pm 2$.
    The non-zero impact parameter in the $x$-direction is
    apparent.
    The shocks move in the $\pm z$ direction at the speed
    of light and collide at $t = z = 0$.
    After the collision the remnants of the initial shocks,
    which remain close to the lightcone,
    $z = \pm t$, are significantly attenuated in amplitude
    with the extracted energy deposited in the interior region.
    The development of transverse flow is apparent
    at positive times.
    \label{fig:snapshots}
    } 
\end{figure*}

\begin{figure}
\begin{center}
\suck[scale = 0.6]{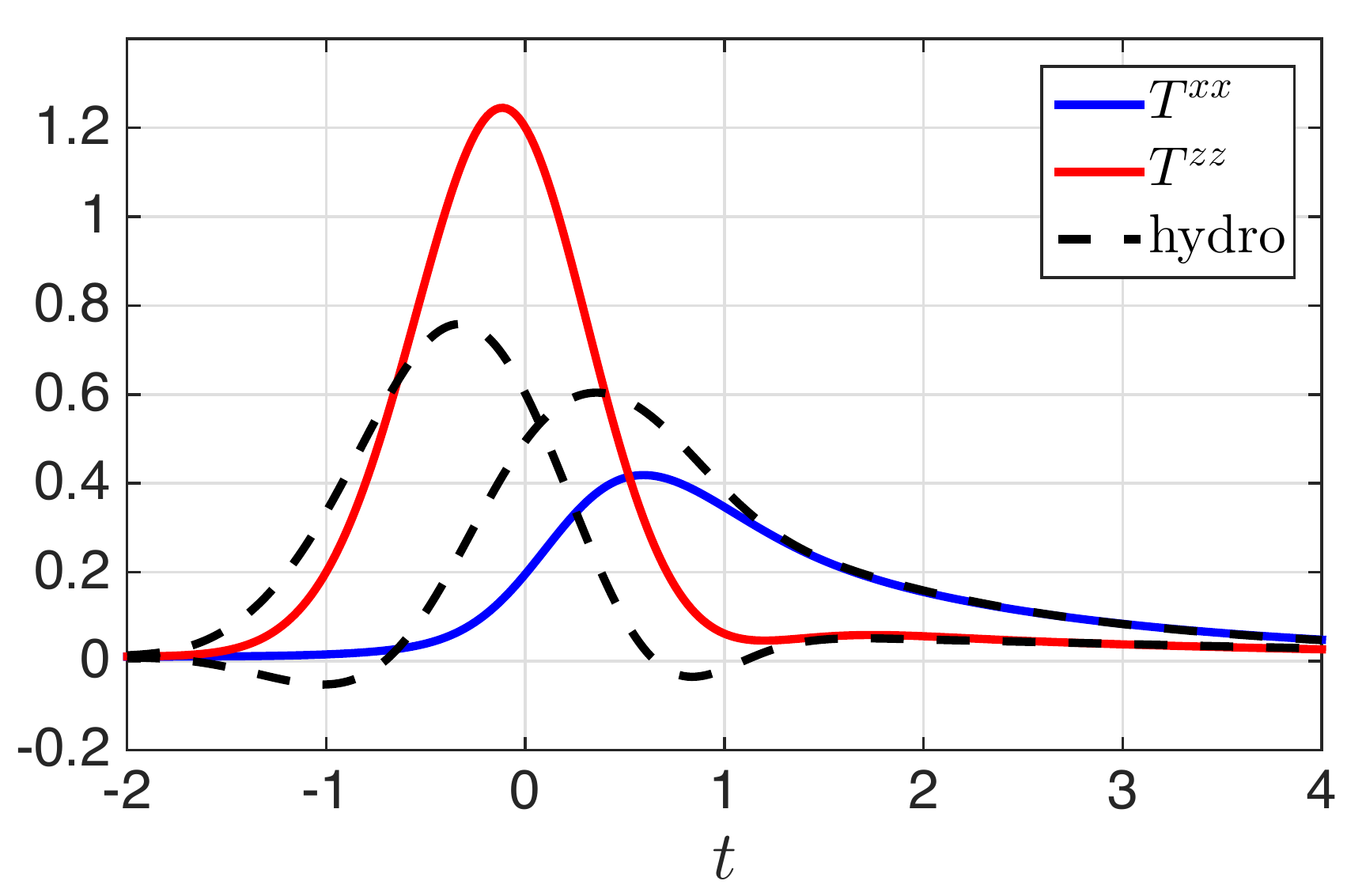}
\end{center}
\caption{Stress tensor components $ T^{xx}$ and $ T^{zz}$ 
at the spatial origin, $x = y = z = 0$, as a function of time.  Dashed lines denote 
the viscous hydrodynamic approximation.  Around $t = 0$ the system is highly anisotropic
and far from equilibrium.  Nevertheless, at this point in space, the system begins to 
evolve hydrodynamically at $t \approx 1.25$.
\label{fig:pressures}
} 
\end{figure}

In Fig.~\ref{fig:snapshots} we plot the energy density $T^{00}$
(top) and energy flux $T^{0i}$ (bottom)
in the plane $y = 0$ at several values of time.%
\footnote
    {%
    Here and henceforth, 
    $T^{\mu\nu}$ is really the expectation value of the SYM
    stress-energy tensor divided by
    $\kappa \equiv \Nc^2/(2 \pi^2)$.
    }
Note that the color scaling varies from plot to plot.
The color scaling in the lower plots denotes $| T^{0i}|$ and the 
flow lines indicate the direction of $T^{0i}$.
At time $t = -2$ the system consists of two well separated lumps 
of energy moving towards each other at the speed of light.
The non-zero impact parameter in the $x$-direction is apparent.
At time $t = 0$, when both shocks are centered at $z = 0$,
the energy density and flux differ by only 8\% and 15\%, respectively,
from a linear superposition of two unmodified shocks.
Nevertheless, the subsequent evolution is very different from
two unmodified outgoing shocks:
the remnants of the initial shocks, which remain close to the lightcone,
$z = \pm t$,
are significantly attenuated in amplitude with the extracted
energy deposited in the interior region.

At sufficiently late times,
in accord with fluid/gravity duality \cite{Bhattacharyya:2008jc,Baier:2007ix},
the evolution of the stress tensor should be governed by hydrodynamics.
To compare to hydrodynamics, we define the fluid velocity 
to be the normalized time-like
$(u_\mu u^\mu = -1)$
future directed
eigenvector of the stress tensor,
\begin{equation}
T^{\mu}_{\ \nu} \, u^\nu = -\epsilon \, u^\mu \,,
\label{eq:veldef}
\end{equation}
with $\epsilon$ the proper energy density.  Given the flow field $u^\mu$
and energy density $\epsilon$,
we construct the hydrodynamic approximation to the stress tensor $T^{\mu \nu}_{\rm hydro}$
using the constitutive relations of first order viscous hydrodynamics.

In Fig.~\ref{fig:pressures} we plot the stress tensor components $ T^{xx}$ and $T^{zz}$,
and their hydrodynamic approximations $T^{xx}_{\rm hydro}$ and $T^{zz}_{\rm hydro}$,
at the spatial origin $x = y = z = 0$, as a function of time.
At this point the stress tensor is diagonal,
the flow velocity $\bm u = 0$,
and $T^{xx}$ and $T^{zz}$ 
are simply the pressures in the $x$ and $z$ directions.  As shown in the figure, 
the pressures increase dramatically during the collision, reflecting a 
system which is highly anisotropic and far from equilibrium.
After a time $t \approx 1.25$
the pressures are well described by viscous hydrodynamics.  Remarkably, at this 
time the transverse pressure $T^{xx}$ is nearly ten times larger than $T^{zz}$.
(This latter phenomena has also been seen in
$1+1$ dimensional flow
\cite{Chesler:2009cy,Chesler:2010bi,Casalderrey-Solana:2013aba}.)

\begin{figure}
\begin{center}
\suck[scale = 0.6]{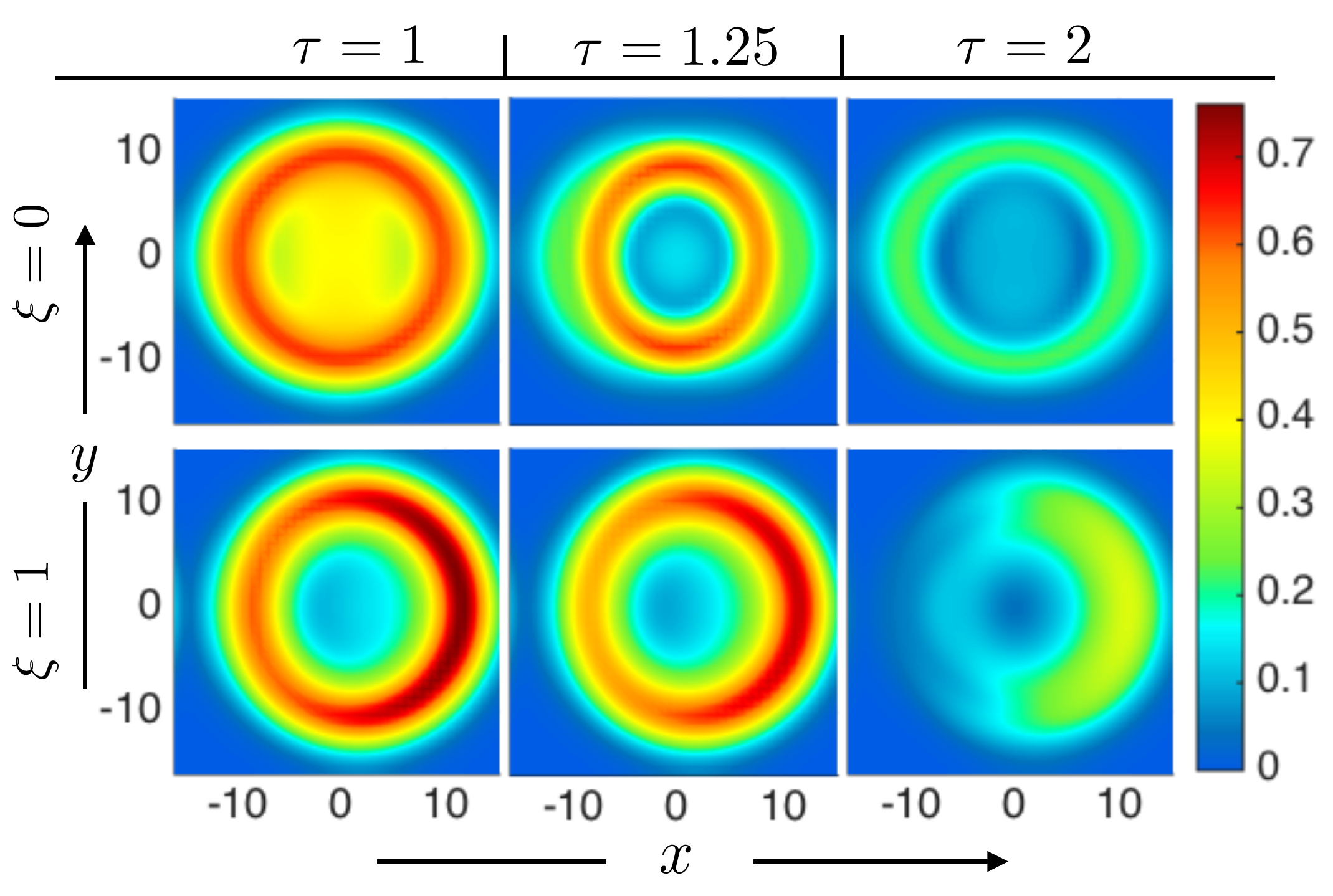}
\end{center}
\caption
    {%
    The residual $\Delta$ in the transverse plane,
    at several proper times $\tau$
    and two values of rapidity $\xi$.
    Regions with $\Delta \ll 1$ are evolving hydrodynamically.  
    At $\xi = 0$ (top row) the central region becomes
    hydrodynamic at $\tau \approx 1.25$,
    whereas
    at $\xi = 1$ (bottom row)
    hydrodynamic behavior of
    the central region has already begun by $\tau \approx 1$.
    At $\xi = 1$, hydrodynamic behavior first sets in at $x < 0$.
    This feature reflects the fact that the receding maxima
    remain far from equilibrium and non-hydrodynamic,
    and 
    the maxima with $\xi > 0$ lies at $x > 0$.
    \label{fig:therm}
    } 
\end{figure}

To quantify the domain in which hydrodynamics is applicable,
we define a residual measure 
\begin{equation}
\label{eq:deltadef}
\Delta \equiv ({1}/{\bar p})\sqrt{\Delta T_{\mu \nu}\Delta T^{\mu \nu}},
\end{equation}
with $\Delta T^{\mu \nu} \equiv T^{\mu \nu} - T^{\mu \nu}_{\rm hydro}$ and 
$\bar p \equiv {\epsilon}/{3}$ the average 
pressure in the local rest frame.
The quantity $\Delta$ is frame-independent but,
when evaluated in the local fluid rest frame,
reduces to
the relative difference between the spatial stress in
$T^{\mu \nu}$ and $T^{\mu \nu}_{\rm hydro}$.
Regions with $\Delta \ll 1$ are evolving hydrodynamically.

In Fig.~\ref{fig:therm} we plot $\Delta$ in the transverse plane 
at proper times $\tau = 1$, 1.25, and 2,
and rapidities $\xi = 0$ and 1.
The color scaling is the same in all plots.
Focusing first on $\xi = 0$ (top row),
at $\tau = 1$ one sees that
$\Delta \gtrsim 0.5$ in the central region ($x,y \approx 0$),
and hydrodynamics is not a good description.
However, by $\tau = 1.25$ a fluid droplet with
$\Delta \lesssim 0.15$ and
transverse radius $x_{\perp} \equiv |\bm x_{\perp}| \lesssim 5.3$
has formed,
with subsequent evolution well described by hydrodynamics.
At $\tau = 2$ the transverse size 
of the droplet has increased and $\Delta < 0.15$ for $x_{\perp}  \lesssim 8.6$.
Turning now to
the behavior at rapidity $\xi = 1$ (bottom row),
one sees that for small $x_{\perp} $
the system is already evolving hydrodynamically
at $\tau = 1$.
Moreover, the onset of hydrodynamics occurs earlier for $x < 0$ than for $x>0$.
This feature reflects the fact that the receding maxima
remain far from equilibrium and non-hydrodynamic,
and (as seen in Fig.~\ref{fig:snapshots}),
the maxima with $\xi > 0$ lies at $x > 0$.

\begin{figure*}[ht]
\suck[scale = 0.5]{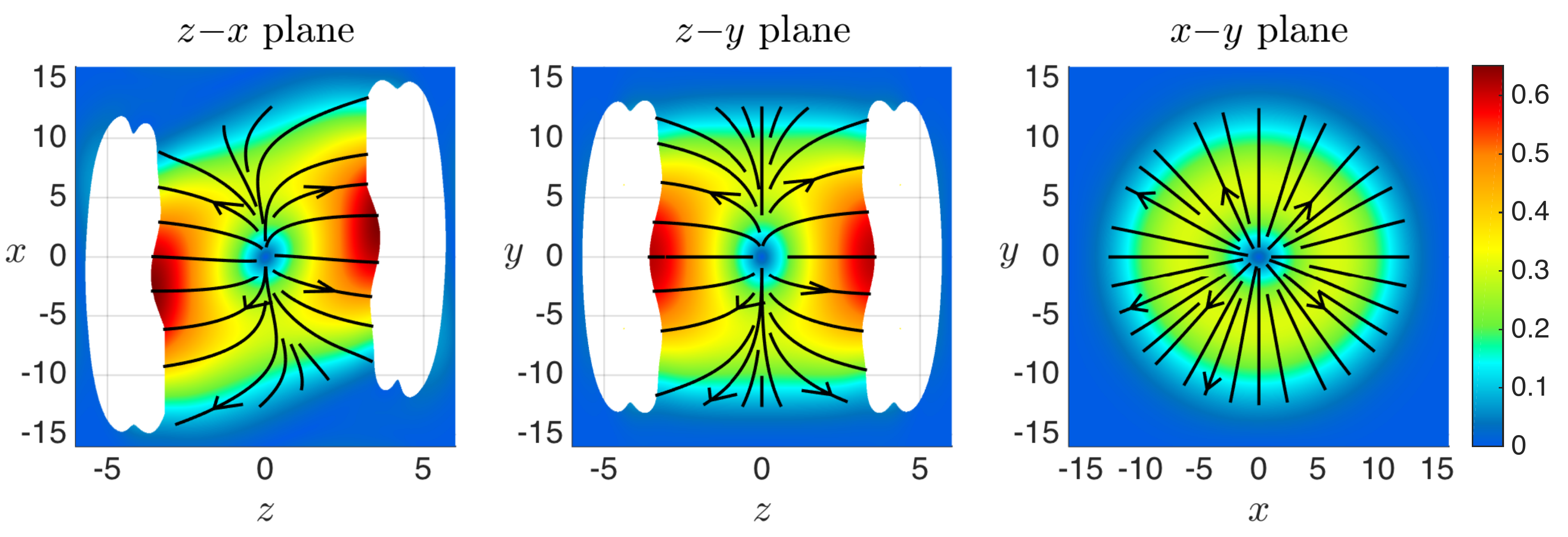}
\caption{The fluid 3-velocity $|\bm v|$ at time $t = 4$, in the $z{-}x$, $z{-}y$,
and $x{-}y$ planes.  
Streamlines denote the direction of $\bm v$.
Regions in which the residual $\Delta$,
defined in Eq.~(\ref{eq:deltadef}), is greater than 0.15 have been excised;
within these regions the system is behaving non-hydrodynamically.
The maximum of $|\bm v|$, which occurs in the vicinity of the receding maxima, is 0.64. 
In contrast, the maximum transverse velocity in the $x{-}y$ plane is 0.3.
\label{fig:velocities}
} 
\end{figure*}

Interestingly, the inclusion of transverse dynamics seems to hasten the approach to local equilibrium: 
the equilibration time $t_{\rm hydro} \sim 1.25$ is about $30\%$ smaller than was the case
in our previous studies \cite{Chesler:2010bi,Chesler:2013lia} of planar shock collisions.
Recent work \cite{Heller:2012km,Heller:2013oxa} has found that equilibration 
time scales of far-from-equilibrium states can be understood,
at least semi-quantitatively, in terms of the spectrum of quasinormal modes.  
Post-collision, a distribution of quasinormal modes will be excited.
With the inclusion of transverse dynamics,
the dominantly excited quasinormal modes will be ones with
non-zero transverse wavevectors of order $1/R$.
Faster decay rates of
quasinormal modes with increasing $|k_\perp|$ should lead to faster
apparent relaxation of the entire sum (i.e., the metric perturbation)
independent of location in the transverse plane.
This will be the case provided one considers sufficiently coarse
measures of relaxation, such as our $\Delta < 0.15$ criterion,
which can become satisfied when numerous quasinormal modes are
comparably excited. 
With a much more stringent local equilibration criterion,
we would expect to see less of a
difference relative to the planar case.

A striking feature of the the post-collision evolution in Fig.~\ref{fig:snapshots} is the appearance 
of flow in the transverse plane at early times.  The early-time acceleration imparted on the 
transverse flow can have a significant impact on the subsequent transverse expansion.
In Fig.~\ref{fig:velocities} we plot the  fluid 3-velocity $\bm v \equiv \bm u/u^0$
in the $z{-}x$, $z{-}y$, and $x{-}y$ planes at time $t = 4$.
The color scaling, which indicates $|\bm v|$, is the same in each plot.
The flow lines show the direction of $\bm v$.
Regions in which $\Delta > 0.15$, and the system is not behaving hydrodynamically,
have been excised.
Already at time $t = 4$ and radius $x_{\perp}  \approx 5$ the transverse fluid velocity in
the $x{-}y$ plane has magnitude 0.3.
In contrast, the maximum of the longitudinal velocity, which occurs in the neighborhood 
of the receding maxima, is 0.64.  

\begin{figure}
\begin{center}
\suck[scale = 0.55]{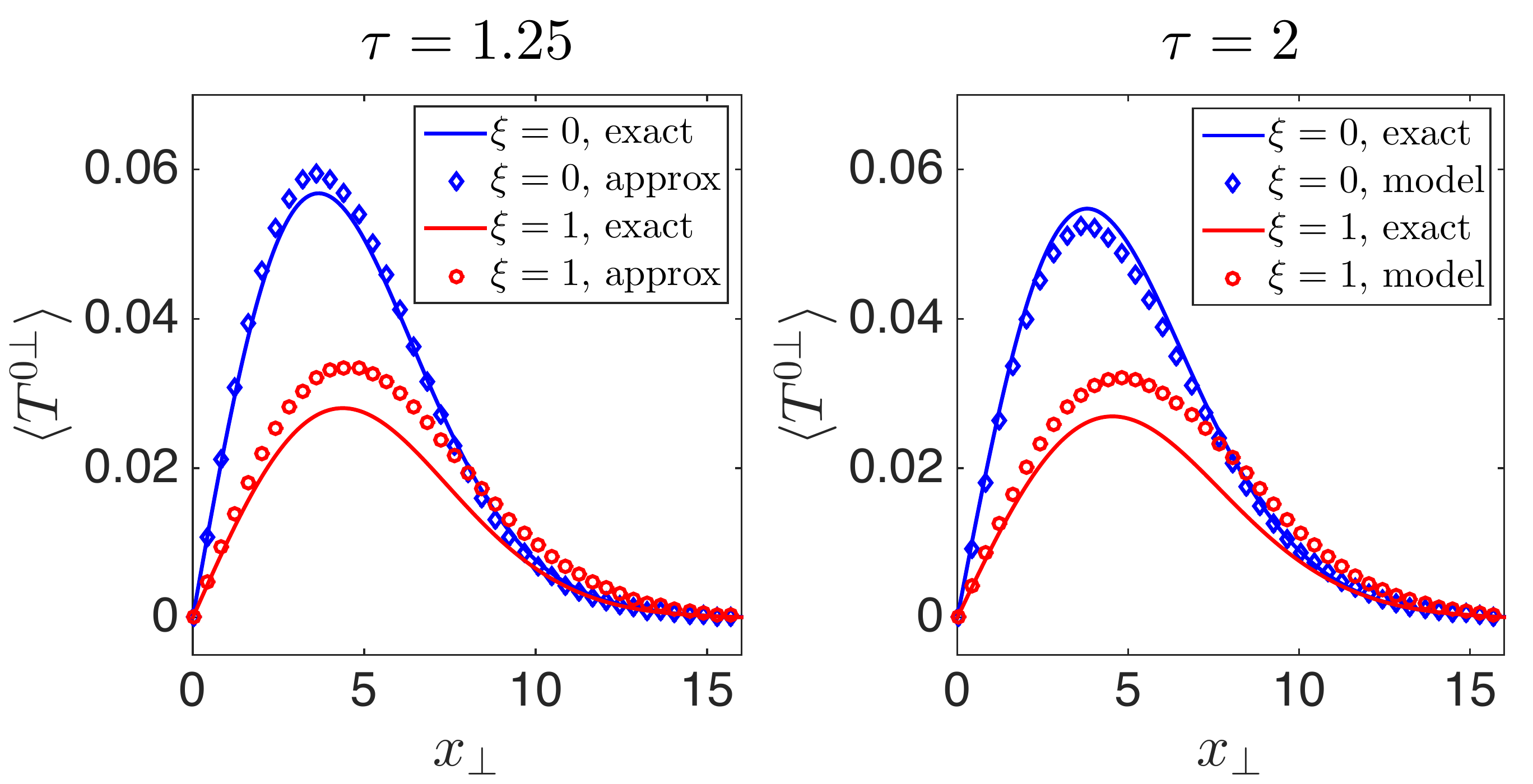}
\end{center}
\vskip -0.15in
\caption{The average lab-frame radial energy flux,
as a function of $x_\perp$, at two values each of 
proper time and rapidity.
Also shown is the approximate form (\ref{eq:pratt})
which, for small rapidity,
agrees quite well with the full results.
\label{fig:radialflow}
} 
\end{figure}

One sees from Fig.~\ref{fig:velocities} that the fluid velocity in the
$x{-}y$ plane is nearly radial: we see no strong signatures of elliptic flow.
At vanishing rapidity (or $z = 0$),
the traceless part of the spatial stress, as well as the
hydrodynamic residual shown in the upper row of fig.~\ref{fig:therm},
deviate significantly from rotational symmetry in the $x{-}y$ plane
(at a 5--6\% level at $t=4$).
However, the energy density and the resulting fluid velocity deviate
only slightly from azimuthal symmetry;
at $t = 4$ their transverse plane anisotropy is about 1\%.
Our computed evolution does not extend through the entire hydrodynamic
phase of the plasma and continued hydrodynamic expansion may generate
more elliptic flow.
However, the very modest elliptic flow,
despite the substantial impact parameter,
may reflect an unphysical aspect of our choice of Gaussian initial
energy density profiles.
Gaussian profiles differ, of course, from more realistic models of
nuclear density (such as Wood-Saxon profiles).
But a Gaussian profile in the transverse plane
is computationally convenient.
The very rapid decrease of its Fourier transform allowed us to
use a rather modest transverse grid.
However, a special feature of Gaussian transverse plane profiles is that
the overlap function (i.e., the product of the energy densities of the
two incident projectiles) is rotationally symmetric for any impact parameter:
\begin{equation}
    e^{-\frac 12(\bm x_\perp + \bm b/2)/R^2} \,
    e^{-\frac 12(\bm x_\perp - \bm b/2)/R^2} \,
    =
    e^{-(\bm x_\perp^2 + (\bm b/2)^2)/R^2} \,.
\end{equation}
With Gaussian profiles (and equal size incident projectiles), there simply
is no ``almond'' in the overlap function.  The actual dynamics depends,
of course, on the full spacetime dependence of the initial data, not
just on this overlap function and, as noted above,
the stress tensor and the residual $\Delta$
depart significantly from transverse rotational symmetry.
Nevertheless, a simple picture in which the early
transverse flow reflects, in large part, this initial
overlap function appears plausible.

A notable feature in our results is that the fluid flow,
in the $z{-}x$ plane, is not symmetric about the $z$ axis
and the longitudinal flow does not vanish at $z = 0$.
The latter observation is a direct violation of the
simplified model of boost invariant flow,
in which $v^z = z/t$ and vanishes at $z =0$.
Nevertheless, at $t = 4$ and in the region of space where $\epsilon > 0.6 \, {\rm max}( \epsilon)$,
the longitudinal flow is roughly described by
boost invariant flow at the 20\% level or better, with larger 
deviations appearing at larger rapidities.%
\footnote
  {%
  Specifically, in the region where $\epsilon > 0.6 \, {\rm max}\,( \epsilon)$,
  we find ${\rm max}\, | v^z - z/t| < 0.20 \> {\rm max}\, | v^z|$.
  }
For planar shock collisions, the deviation of the longitudinal fluid velocity 
from boost invariant flow decreases as the shock thickness decreases
\cite{Chesler:2013lia,Casalderrey-Solana:2013aba}.%
\footnote
  {%
  However, even in the limit of thin planar shocks
  the proper energy density has strong rapidity dependence
  \cite{Casalderrey-Solana:2013aba,Chesler:2013lia}.
  }
It will be interesting to see if this also holds when  
transverse dynamics is included.

We conclude by discussing the early-time transverse flow predicted in
ref.~\cite{Vredevoogd:2008id}.
There, using assumptions of boost invariance and transverse plane rotational symmetry,
it is argued that at early times the transverse energy flux
is proportional to the gradient of the energy density and
grows linearly with time,
\begin{equation}
\label{eq:pratt}
T^{0x} = -{\textstyle \frac{t}{2}} \, \partial_x \epsilon \,,  \qquad
T^{0y} = -{\textstyle \frac{t}{2}} \, \partial_y \epsilon \,.
\end{equation}
In Fig.~\ref{fig:radialflow} we plot the angular averaged radial flow
$\langle T^{0\perp} \rangle \equiv \langle \hat x_{\perp}^i T^{0i} \rangle$,
together with the approximation (\ref{eq:pratt}),
at proper times $\tau = 1.25$ and 2, and rapidities $\xi = 0$ and 1.
The approximation (\ref{eq:pratt}) works remarkably well at both times and 
rapidities, although the agreement is not quite as good at $\xi = 1$ where
the assumption of boost invariance is more strongly violated.  
It would be interesting to see if the agreement with the approximation (\ref{eq:pratt}) 
improves when the shock thickness decreases.

\section {Final remarks}

We have presented results from the first calculation,
using numerical holography, of the evolution in strongly
coupled $\Nfour$ SYM theory of an initial state which
resembles two colliding heavy nuclei,
Correctly treating the dynamics, both longitudinal
and transverse, requires solving a five dimensional
gravitational initial value problem.
This was challenging, but feasible, using our
characteristic formulation and (good)
desktop-scale computing resources.
The results presented above are a proof-of-principle.
Clearly, it will be interesting to explore how results
change as the impact parameter, longitudinal thickness,
and transverse size of each projectile are varied.
It should also be possible to explore the effects produced
by using more realistic initial energy density profiles,
possibly including shorter distance fluctuations 
of the type which have been suggested to be
relevant in real heavy ion collisions.
(See, for example, ref.~\cite{Luzum:2013yya,
Floerchinger:2014fta,Schenke:2014tga,Shen:2015qta} and references therein.)
We hope future work will shed light on some of these topics.

\begin{acknowledgments}
We thank Wilke van der Schee for discussions of prior work. 
The work of PC is supported
by the Fundamental Laws Initiative of the Center for the
Fundamental Laws of Nature at Harvard University.
The work of LY was supported, in part, by the U.S. Department
of Energy under Grant No.~DE-SC0011637.
He thanks the Alexander von Humboldt Foundation
and the University of Regensburg
for their generous support and hospitality during portions of this work.
\end{acknowledgments}

\bibliographystyle{utphys}
\bibliography{refs}%
\end{document}